%% file: paper.tex
\documentclass[prc,twocolumn,showpacs,nofootinbib]{revtex4}
\usepackage{epsfig,amsmath,pictex,graphicx,bm}
\newcommand{\data}[3]{\putbar breadth <5mm> from #1 #2 to #1 #3 }

\newcommand{\calc}[2]{\multiput{\line(1,0){5}} at #1 #2 *1 0 2 / }

\newcommand{\SC}{\scriptstyle}

\newcommand{\mev}{\mbox{Me\hspace{-0.1em}V}}
\newcommand{\gev}{\mbox{Ge\hspace{-0.1em}V}}
\newcommand{\einheit}{\bm 1}
\begin{document}

\title{Vector mesons in a relativistic point-form approach}

\author{A.~Krassnigg and W. Schweiger}
\affiliation{Institute for Theoretical Physics, University
of Graz,  A-8010 Graz, Austria }
\author{W.~H.~Klink}
\affiliation{Department of Physics and Astronomy, University of
Iowa, Iowa City, IA 52242, USA}
\date{\today}

\begin{abstract}
We apply the point form of relativistic quantum mechanics to develop a
Poincar\'{e} invariant coupled-channel formalism for two-particle
systems interacting via one-particle exchange.  This approach takes
the exchange particle explicitly into account and leads to a
generalized eigenvalue equation for the Bakamjian-Thomas type mass
operator of the system.  The coupling of the exchange particle is
derived from quantum field theory.  As an illustrative example we
consider vector mesons within the chiral constituent quark model in
which the hyperfine interaction between the confined quark-antiquark
pair is generated by Goldstone-boson exchange.  We study the effect of
retardation in the Goldstone-boson exchange by comparing with the
commonly used instantaneous approximation.  As a nice physical feature
we find that the problem of a too large $\rho$-$\omega$ splitting can
nearly be avoided by taking the dynamics of the exchange meson
explicitly into account.
\end{abstract}


\maketitle

\section{Introduction}
In 1949 Dirac formulated a way of incorporating relativity into
quantum theory that differed from quantum field theory \cite{Di49}.
Although Dirac's paper was written in the context of
classical mechanics, his methodology -- the use of representations
of the symmetry group of the theory of special relativity, the
Poincar\'{e} group -- was also applicable to quantum theory \emph{and}
to quantum field theory (for a review, see Ref.~\cite{KePo91}). In
Dirac's original presentation he made evident how to add interactions
to a theory of free particles in agreement with the Poincar\'{e}
algebra, ending up with conditions for the interaction terms that are
in general nonlinear. In 1953, Bakamjian and Thomas gave a
prescription for an explicit construction involving only linear
constraints \cite{BaTh53}.

Of the various forms that Dirac introduced the instant form is the
most widely used \cite{Co65,CoOs75,Sc62,GlKr97}, although almost
exclusively in the context of quantum field theory.  The front form of
Hamiltonian dynamics became popular as a natural framework for
treating parton phenomena.  For a topical review, see
Ref.~\cite{Bretal98}.  The point form of relativistic dynamics has
also been consi\-dered in quantum field theory
\cite{Fuetal73,Gretal74,So74}, but because of its complicated
quantization surface it was not further developed.  Only recently has
the point form been rediscovered, this time in the context of the
quantum mechanics for finite degree-of-freedom systems.  Lev has
analyzed electromagnetic current operators in the point form and then
transformed them to the other two forms \cite{Le95}.  Klink has
suggested a basis of states, called velocity states \cite{Kl98b},
which are suitable for few-body point-form quantum theory.  These
states have also been introduced by Karmanov in a different context
\cite{Ka98,Caetal98}.  Klink made use of velocity states when treating
nuclear physics problems \cite{Kl98a,KlRo98,AlKl98}.  Recently, also
the Graz group employed the point form of relativistic dynamics to
describe the electroweak structure of baryons within a chiral
constituent quark model~\cite{Gletal01,Waetal01,Boetal02}.

This paper uses the point form to elaborate on a coupled-channel
formalism which is applicable to a wide range of problems. As a
first and simple application we have chosen a two-particle system
of one (constituent) quark and one (constituent) antiquark which
form vector mesons.  The hyperfine interaction in this simple
system comes from the chiral constituent quark model with
pseudoscalar meson exchange.  Such an interaction has been used
in a semirelativistic form with great success for the calculation
of baryon spectra~\cite{Gletal98b,Waetal99}.  For vector
mesons~\cite{Th98} the results within this semirelativistic
approach are not as good. In the present paper a fully
relativistic calculation with the exchange-meson channel
explicitly included is presented and compared to the
semirelativistic approach in which the meson-exchange is treated
in an instantaneous approximation.

The necessary formalism is introduced in detail in
Secs.~\ref{pfformalism}--\ref{eigequ}.  In Sec.~\ref{pfformalism}
we summarize relevant features of the Poincar\'{e} group.
Velocity states are introduced as a suitable basis for the
quantum-mechanical treatment of few-particle systems in
point-form.  These states are subsequently used to construct the
elementary meson-(anti)quark vertex, which enters the
invariant-mass operator.  The mass operator is treated in
Sec.~\ref{massops}; in our example its interacting part arises
from a pseudoscalar Hamiltonian density.  Our mass operator is of
Bakamjian-Thomas type and acts on a Hilbert space which is the
direct sum of two-particle and two-plus-one particle Hilbert
spaces. Such an ansatz deals with effective degrees of freedom in
contrast to a quantum field theory.  It leads to the coupled
two-channel problem outlined in Sec.~\ref{qfttoqm}.  The validity
of Poincar\'{e} invariance for systems with a finite (but not
necessarily conserved) number of particles is guaranteed by the
Bakamjian-Thomas construction as described in
Sec.~\ref{qfttoqm}.  The fact that quarks and antiquarks are
always confined is accounted for by adding harmonic oscillator
confinement terms to the square of the kinetic terms of the
coupled-channel mass o\-perator.  As a first step to solve the
eigenvalue problem for the mass operator the two-channel problem
is reduced to a one-channel problem with an optical potential
which depends on the mass eigenvalue to be determined.  The
harmonic-oscillator eigenfunctions of the pure confinement
problem are then used as a basis for expanding the
quark-antiquark wave functions of the full problem including the
hyperfine interaction.  As a result, one can discretize the
dynamical two-particle equation and obtain a set of coupled
algebraic equations that can be solved.  The eigenvalues of the
mass operator are determined by a resonance condition, which takes
into account the nonlinear appearance of eigenvalues in the
eigenvalue equation.  The structure of the equation and the
implementation of confinement are discussed in Sec.~\ref{eigequ}.

Sec.~\ref{numerics} contains some remarks on the numerics and the
specific solution method employed in the calculation.  Comments
on the instantaneous approximation and the choice of the model
parameters are given in Secs.~\ref{nonpot} and~\ref{model},
respectively.  The results of the calculations are vector-meson
masses and some branching ratios of their hadronic decay widths.
The numbers are presented in Sec.~\ref{results}; they are
compared with experimental numbers as well as with the
instantaneous approximation to elucidate retardation effects
coming from the hyperfine-interaction. Concluding remarks
can be found in Sec.~\ref{concl}.  Our conventions,
normalizations and matrix elements required for the calculations
are summarized in the appendix.

\section{Poincar\'{e} Group}
\label{pfformalism}
The starting point for dealing with few-body systems in relativistic
quantum mechanics is the set of commutation relations of the
Poincar\'{e} generators
\begin{eqnarray}\label{covcoma}
\left[P_\mu,P_\nu\right]&=&0\;,\\ \label{covcomb}
\left[J_{\mu\nu},P_\kappa\right]&=&i(g_{\nu\kappa}P_\mu-
g_{\mu\kappa}P_\nu)\;,\\ \nonumber
\left[J_{\mu\nu},J_{\kappa\lambda}\right]&=&-i(g_{\mu\kappa}
J_{\nu\lambda}-g_{\nu\kappa}J_{\mu\lambda}+\\ \label{covcomc}
&&g_{\nu\lambda}
J_{\mu\kappa}-g_{\mu\lambda}J_{\nu\kappa})\;.
\end{eqnarray}

One can write these relations in a global way by defining
$U_\Lambda$ as the unitary operator
representing the Lorentz transformation $\Lambda$ on the Hilbert
space. In the point form all interactions are contained in the
four-momentum operator, so the significant
commutation relation is
\begin{equation}\label{pfcomm1}
[P^\mu,P^\nu]=0\;,
\end{equation}
which states that the components of the four-momentum commute among
each other. The other commutation relation involving $P^\mu$ is
written as
\begin{equation}\label{pfcomm2}
U_\Lambda P^\mu U^{-1}_\Lambda=(\Lambda^{-1})^\mu{}_\nu P^\nu\;,
\end{equation}
which means that the four-momentum operator has to transform as a
four-vector under Lorentz transformations. The commutation
relations of the Lorentz generators among themselves are
unaffected by interactions in the point form. We will refer to
Eqs.~(\ref{pfcomm1}) and (\ref{pfcomm2}) later, when we construct
mass operators.

Let us start with single-particle states $|p,\sigma\rangle$, e.\,g.~for
spin-$\frac{1}{2}$ particles, which transfer irreducibly under the
Poincar\'{e} group. Defining the action of the four-momen\-tum 
operator $P^\mu$ on such single-particle states by
\begin{equation}\label{momdef}
P^\mu|p,\sigma\rangle=|p,\sigma\rangle p^\mu
\end{equation}
one can easily show that Eqs.~(\ref{pfcomm1}) and (\ref{pfcomm2}) are
satisfied for a single-particle representation.  In the following we
need also the Poincar\'{e} transformation properties of such states:
\begin{eqnarray}
U_b|p,\sigma\rangle&=&e^{-ibp}|p,\sigma\rangle\qquad\mbox{and}\\
\label{trawig}
U_\Lambda|p,\sigma\rangle&=&\sum_{\sigma'=\pm\frac{1}{2}}|\Lambda p,
\sigma'\rangle
D^{\frac{1}{2}}_{\sigma'\sigma}(R_W(p,\Lambda))\;.
\end{eqnarray}
$U_b$ denotes a space-time translation by a constant four-vector $b$
and the Wigner rotation $R_W(p,\Lambda)$ is
given by
\begin{equation}\label{wigrot}
R_W(p,\Lambda)=B^{-1}(\Lambda \frac{p}{m})\Lambda B(\frac{p}{m})\;.
\end{equation}
The $D^{\frac{1}{2}}_{\sigma'\sigma}$ are the matrix elements of
the standard Wigner $D$-functions \cite{Vaetal88}.

With Eqs.~(\ref{momdef}-\ref{wigrot}) we see that
Eq.~(\ref{pfcomm1}) is immediately
satisfied. For Eq.~(\ref{pfcomm2}) we have, applying its left-hand
side
to $|p,\sigma\rangle$,
\begin{eqnarray}\nonumber
&&U_\Lambda P^\mu U^{-1}_\Lambda|p,\sigma\rangle=
\\\nonumber
&=&U_\Lambda P^\mu U_{\Lambda^{-1}}|p,\sigma\rangle
\\\nonumber
&=&U_\Lambda P^\mu\!\sum_{\sigma'}|
\Lambda^{-1} p,\sigma'\rangle D^{\frac{1}{2}}_{\sigma'\sigma}
(R_W(p,\Lambda^{-1}))
\\\nonumber
&=&U_\Lambda\sum_{\sigma'}|\Lambda^{-1} p,\sigma'\rangle
(\Lambda^{-1})^\mu{}_\nu p^\nu
D^{\frac{1}{2}}_{\sigma'\sigma}(R_W(p,
\Lambda^{-1}))
\\\nonumber
&=&U_\Lambda U_{\Lambda^{-1}}|p,\sigma\rangle (\Lambda^{-1}
)^\mu{}_\nu p^\nu
\\\nonumber
&=&U_{\Lambda\Lambda^{-1}}|p,\sigma\rangle (\Lambda^{-1}
)^\mu{}_\nu p^\nu
\\
&=&(\Lambda^{-1})^\mu{}_\nu P^\nu|p,\sigma\rangle\;,
\end{eqnarray}
which is the desired expression for the right-hand side.
In this derivation we have extensively used the representation
properties of $U_\Lambda$ and $U^{-1}_\Lambda$, e.\,g.~that
\begin{equation}
U_\Lambda U_{\Lambda^{-1}}=U_{\Lambda\Lambda^{-1}}=U_{1}=
\einheit \;.
\end{equation}

In the generalization of single-particle states to multiparticle states, it
is useful to introduce velocity sta\-tes, which have simple
transformation properties under Lorentz transformations.
We start with usual multiparticle momentum states which are tensor products
of irreducible representations of the Poincar\'{e} group.
We observe that under a Lorentz-trans\-formation (see Eq.~(\ref{trawig}))
\begin{widetext}
\begin{eqnarray}
U_\Lambda |p_1,\sigma_1,p_2,\sigma_2,\ldots,p_n,\sigma_n\rangle=
\sum_{\sigma_1',\sigma_2',\ldots,\sigma_n'=\pm\frac{1}{2}}
|\Lambda p_1,\sigma_1',\Lambda p_2,\sigma_2',\ldots,\Lambda p_n,
\sigma_n'\rangle\;\prod_{i=1}^n
D^{\frac{1}{2}}_{\sigma_i'\sigma_i}(R_{W}(p_i,\Lambda))\;,
\end{eqnarray}
\end{widetext}
where each of the $D$-functions depends on a \emph{different}
Wigner rotation $R_{W}(p_i,\Lambda)$. This implies that one
cannot couple angular momenta in the standard way. So it is
desirable to have more uniform transformation properties of
$n$-particle states under a Lorentz transformation. This is the
case for velocity states. For such states all spin projections and
also the individual particle momenta are subject to the {\em
same} Wigner rotation and the effect of the Lorentz
transformation goes mainly into the overall velocity $v$. We now
show the construction of velocity states in detail and make their
Lorentz transformation properties evident.

We consider an $n$-particle system with particle momenta
$p_i$ and spin projections $\sigma_i$, $i=1,\dots,n$, and start by
defining internal momenta $k_i$ via
\begin{equation}\label{pointboost}
k_i=B^{-1}_c(v)p_i\;,
\end{equation}
where $B_c(v)$ is a canonical spin boost, i.\,e.~a rotationless Lorentz
transformation, which transforms our system from its rest frame to
total velocity $\vec{v}$.
%
%
%
%
The momenta $\vec{k}_1,\vec{k}_2,\ldots,\vec{k}_n$ satisfy
\begin{equation}\label{momsumzero}
\sum_{i=1}^n \vec{k}_i=0
\end{equation}
and thus only $n-1$ of them are linearly independent. The
remaining independent variable is the overall four velocity $v$
of the system.

The construction of a velocity state can be viewed as starting from a
multiparticle momentum state in its rest frame.  This state is then
boosted to overall velocity $v$ by means of the canonical spin boost
whose inverse is used in Eq.~(\ref{pointboost}) to yield the velocity
state
\begin{widetext}
\begin{eqnarray}\label{velstate}
&&|v,\vec{k}_1,\mu_1,\vec{k}_2,\mu_2,\ldots,\vec{k}_n,\mu_n\rangle:=
U_{B_c(v)}|k_1,\mu_1,k_2,\mu_2,\ldots,k_n,\mu_n\rangle\\
\label{velstate1}
&&=\sum_{\sigma_1,\sigma_2,\ldots,\sigma_n=\pm\frac{1}{2}}
|p_1,\sigma_1,p_2,\sigma_2,\ldots,p_n,
\sigma_n\rangle\;\prod_{i=1}^n
D^{\frac{1}{2}}_{\sigma_i\mu_i}\left(R_{W}(k_i,B_c(v))\right)
\;.\qquad
\end{eqnarray}
\end{widetext}
This equation makes evident that a velocity state is a linear
combination of multiparticle momentum states. We also note that
velocity states transform irreducibly under transformations of the
Poincar\'{e} group. Concerning notation we in general write
$\sigma_i$ to denote spin projection variables, but for velocity
states and when using internal variables of a system we will
instead write $\mu_i$ for spin projections to make a clear
distinction between general and internal variables. In this sense
the $\sigma_i$ should always appear together with the $p_i$,
whereas the $\mu_i$ appear together with the $\vec{k}_i$.

Next we study the Lorentz-transformation properties of a
velocity state. We again apply the general boost operator
$U_\Lambda$ to the velocity state~(\ref{velstate}) and combine its
action with that of $U_{B_c(v)}$. Using Eq.~(\ref{wigrot}) one
obtains

\begin{eqnarray}\nonumber
&&U_\Lambda|v,\vec{k}_1,\mu_1,\vec{k}_2,\mu_2,\ldots,\vec{k}_n,\mu_n
\rangle=\\ \nonumber
&=&U_\Lambda U_{B_c(v)}|k_1,\mu_1,k_2,\mu_2,\ldots,k_n,\mu_n\rangle
\\ \nonumber
&=&U_{\SC\Lambda B_c(v)}
|k_1,\mu_1,k_2,\mu_2,
\ldots,k_n,\mu_n\rangle\\ \nonumber
&=&U_{B_c(\Lambda v)\,R_W}|k_1,\mu_1,k_2,\mu_2,
\ldots,k_n,\mu_n\rangle\\
&=&U_{B_c(\Lambda v)}U_{R_W}|k_1,\mu_1,k_2,\mu_2,
\ldots,k_n,\mu_n\rangle
\end{eqnarray}
with the Wigner rotation
\begin{equation}
R_W=B_c^{-1}(\Lambda v)\Lambda B_c(v)\;.
\end{equation}
Evaluating the action of $U_{R_W}$ on the velocity state, we get
\begin{widetext}
\begin{eqnarray}\nonumber
U_\Lambda|v,\vec{k}_1,\mu_1,\vec{k}_2,\mu_2,\ldots,\vec{k}_n,\mu_n
\rangle&=&U_{B_c(\Lambda v)}U_{R_W}|k_1,\mu_1,k_2,\mu_2,
\ldots,k_n,\mu_n\rangle\\ \nonumber
&=&U_{B_c(\Lambda
v)}\sum_{\mu_1',\mu_2',\ldots,\mu_n'=\pm\frac{1}{2}}
|R_W k_1,\mu_1',R_W k_2,\mu_2',\ldots,R_W k_n,\mu_n'\rangle
\prod_{i=1}^{n}D^{\frac{1}{2}}_{\mu_i'\mu_i}(R_{W})\\
\label{veltrans}
&=&\sum_{\mu_1',\mu_2',\ldots,\mu_n'=\pm\frac{1}{2}}
|\Lambda v,R_W \vec{k}_1,\mu_1',R_W \vec{k}_2,\mu_2',\ldots,R_W
\vec{k}_n,\mu_n'\rangle \prod_{i=1}^{n}
D^{\frac{1}{2}}_{\mu_i'\mu_i}(R_{W})\;.
\end{eqnarray}
\end{widetext}
In this derivation use has been made of the fact that for
canonical spin boosts the Wigner rotation $R_W$ corresponding to a
rotation $R$ is the rotation itself. It is helpful to notice here
that a rotation is also a Lorentz transformation. One can now
clearly see that the rotation appearing in the $D$-functions and
in the state is the same for all $\mu_i$ and all $\vec{k}_i$. So
one can couple spins and also orbital angular momenta using the
standard addition rules, which was the desired goal.

\section{Mass Operator}
\label{massops} In this section we discuss properties of the mass
operator and its role in our approach.

In Refs.~\cite{BaTh53,KePo91} one can find a general procedure for
adding interactions to a system of free relativistic particles so
that Poincar\'{e} invariance is preserved.  Such a procedure,
called the Bakamjian-Thomas construction, adds an interaction to
the free mass operator.  The Bakamjian-Thomas construction in the
point form involves the (free) four velocity operator $V^\mu_0$
which is introduced by expressing the free four-momentum operator
as
\begin{equation}\label{freefourmom}
P^\mu_0=M_0V^\mu_0\;.
\end{equation}

Interactions are added by perturbing the free mass operator,
$M_0\rightarrow M=M_0+M_I$, in such a way that
Eqs.~(\ref{pfcomm1},\ref{pfcomm2}) are satisfied (i.\,e.~the components
of the four momentum must commute with each other and they have
to transform as the components of a four vector under Lorentz
transformations). We emphasize once more that this formulation
reflects the fact that interactions do not enter the Lorentz
generators, but solely the components of the four momentum. The
linear constraints on the interacting part of the mass operator
are that it should be a Lorentz scalar and commute with the free
four velocity. The interacting four momentum operator is then
reconstructed by
\begin{equation}\label{intfourmom}
P^\mu=M\;V^\mu_0\;,
\end{equation}
where $M$ contains the interactions and $V^\mu_0$ is still
kinematical.

Before starting to construct explicitly an interacting mass
o\-perator, we will discuss the free four-momentum and mass
operators. Let us first examine the effects of the free
four-momentum operator $P_0^\mu$, $\mu = 0,1,2,3$, on the
velocity states defined in Sec.~\ref{pfformalism}. We consider a
system of $n$ free particles with masses $m_i$, internal momenta
$\vec{k}_i$, overall four velocity $v$ and spin projections
$\mu_i$. Then we define
\begin{equation}\label{statemass}
\mathcal{M}_n:=\sum_{i=1}^n \sqrt{m_i^2+\vec{k}_i^2}
\end{equation}
which is the free relativistic mass of the system. Recalling the
action of the free four-momentum operator $P_0^\mu$ on usual
n-particle momentum states, using equations (\ref{velstate}),
(\ref{pfcomm2}) and (\ref{statemass}), and evaluating the boost
explicitly we get
\begin{eqnarray}\nonumber
&&P^\mu_{0}|v,k_1,\mu_1,k_2,\mu_2,\ldots,k_n,\mu_n\rangle=\\\nonumber
&&=\!\begin{pmatrix}\mathcal{M}_n\sqrt{1+\vec{v}^2}\\
\mathcal{M}_n\vec{v}\end{pmatrix}\!
|v,\vec{k}_1,\mu_1,\vec{k}_2,\mu_2,\ldots,\vec{k}_n,\mu_n\rangle
\\&&=
\mathcal{M}_nv^\mu|v,\vec{k}_1,\mu_1,\vec{k}_2,\mu_2,\ldots,
\vec{k}_n,\mu_n\rangle\;.
\end{eqnarray}
Hence a velocity state
$|v,\vec{k}_1,\mu_1,\vec{k}_2,\mu_2,\ldots,\vec{k}_n,\mu_n\rangle$ is
an eigenstate of $P^\mu_{0}$ with the eigenvalue $\mathcal{M}_nv^\mu$. 
Thus we can split $P^\mu_{0}$ according to Eq.~(\ref{freefourmom}) and
a velocity state
$|v,\vec{k}_1,\mu_1,\vec{k}_2,\mu_2,\ldots,\vec{k}_n,\mu_n\rangle $
becomes also an eigenstate of $M_{0}$ and $V^\mu_{0}$ with the
eigenvalues $\mathcal{M}_n$ and $v^\mu$, respectively:
\begin{eqnarray}\label{freemassop}
M_{0}\;|v,\vec{k}_1,\mu_1,\ldots,\vec{k}_n,\mu_n\rangle&=&
\mathcal{M}_n\;|\vec{k}_1,\mu_1,\ldots,\vec{k}_n,\mu_n\rangle\nonumber \\
V^\mu_{0}\;|v,\vec{k}_1,\mu_1,\ldots,\vec{k}_n,\mu_n\rangle&=&
v^\mu\;|\vec{k}_1,\mu_1,\ldots,\vec{k}_n,\mu_n\rangle
\nonumber \\ \phantom{haha}
\end{eqnarray}
The decomposition of Eq.~(\ref{freefourmom}) is the starting
point of the Bakamjian-Thomas construction in point form. This
result is also consistent with the common definition of the mass
operator
\begin{equation}
M_{0}^2=P^\mu_{0}P_{0\mu}\;,
\end{equation}
since the square of the four velocity is always the identity.

\section{Bakamjian-Thomas type vertex interaction}
\label{qfttoqm} In this section we will show how an interacting mass
operator that couples $i$ and $(i+1)$ particle channels can be derived
from a field theoretical vertex interaction such that it fits into the
Bakamjian-Thomas framework.  Our presentation takes up the procedure
suggested in Ref.~\cite{Kli03} to which we also refer for further
details.  We will set up a model with a finite number of effective
degrees of freedom and consider the dynamical equation that describes
this system of interacting particles.

On a direct-sum Hilbert space for $i$ and $i+1$ particles the
(full interacting) mass operator $M$ becomes a matrix operator
\begin{equation}\label{ccmassop}
M= M_0+M_I=
\begin{pmatrix}\mathcal{D}_i^0&0\\0&\mathcal{D}_{i+1}^0\end{pmatrix}
+\begin{pmatrix}0&K^\dagger\\K&0\end{pmatrix}\;.
\end{equation}
The two subspaces are coupled by the vertex operator $K$ and
$\mathcal{D}_i^0$ denotes a free (indicated by the superscript
${}^0$) $i$-particle operator, which corresponds to the
relativistic $i$-particle mass~(\ref{statemass}). In order
to obtain an expression for the vertex operator $K$ we consider a
field theoretical Hamiltonian density $\mathcal{H}_I(x)$, which
describes a vertex interaction and is a polynomial in free
fields, meaning that
\begin{equation}
U_\Lambda\mathcal{H}_I(x)U_\Lambda^{-1}=\mathcal{H}_I(\Lambda x)\;.
\end{equation}
Then $\mathcal{H}_I(0)$ is a Lorentz scalar, since
\begin{equation}
U_\Lambda\mathcal{H}_I(0)U_\Lambda^{-1}=\mathcal{H}_I(0)\;.
\end{equation}
These properties of $\mathcal{H}_I(0)$ can be used to define $K$.
Taking the velocity-state representation and keeping in mind that
the whole velocity dependence of a Bakamjian-Thomas type mass
operator in point form is merely a factor $\propto v_{0} \delta^3
(\vec{v}^\prime -\vec{v})$ (see App.~\ref{defs}) we are led to
introduce $K$ via the relation
\begin{widetext}
\begin{eqnarray}\nonumber
&&\langle
v,\vec{k}_1,\mu_1,\vec{k}_2,\mu_2,\ldots,\vec{k}_{i+1},\mu_{i+1}|
K|v',\vec{k}_1',\mu_1',\vec{k}_2',\mu_2',\ldots,
\vec{k}_{i}',\mu_{i}'\rangle\propto\\ \label{meintmass} &&
v_0\;\delta^3(\vec{v}-\vec{v}')f(\Delta m) \langle
v=0,\vec{k}_1,\mu_1,\vec{k}_2,\mu_2,\ldots,\vec{k}_{i+1},\mu_{i+1}|
\mathcal{H}_I(0)|v=0,\vec{k}_1',\mu_1',\vec{k}_2',\mu_2',\ldots,
\vec{k}_{i}',\mu_{i}'\rangle\;.\qquad \label{intmassop}
\end{eqnarray}
\end{widetext}
The matrix element on the right-hand side of
Eq.~(\ref{intmassop}) has to be understood such that two
particles on the ``left'' and one on the ``right'' are coupled by
$\mathcal{H}_I(0)$ in all possible ways. The remaining particles
yield spectator conditions. This means that $M_{I}$ is constructed
from matrix elements of $\mathcal{H}_{I}(0)$ between velocity
states with the {\em same overall velocity} which furthermore can
be taken to be zero since $\mathcal{H}_{I}(0)$ is a Lorentz
scalar.  What we have neglected in this kind of construction as
compared to the full interacting field theory with vertex
interaction $\mathcal{H}_{I}$ are off-diagonal terms in the
overall velocity, which should not occur in a
Bakamjian-Thomas-type mass operator. Part of such terms can be
simulated with an appropriate choice of the vertex form factor
$f(\Delta m)$ which guarantees also that the mass operator is a
well defined operator on the Hilbert space.  For velocity state
matrix elements ($v'=v$) this form factor can be expressed as
\begin{equation}
    f[(p'-p)^2]=f[(\mathcal{M}_{i+1}'v'-\mathcal{M}_i v)^2]=f[(\Delta m)^2] \,
\label{formf}
\end{equation}
since $v^2=1$.

For our subsequent application to vector mesons the matrix
element~(\ref{intmassop}) should describe the coupling of a
pseudoscalar meson to a quark. It has the particular form
\begin{widetext}
\begin{eqnarray}\nonumber
&&\langle v,\vec{k}_1,\mu_1,\vec{k}_2,\mu_2,\vec{k}_3|K|
v',\vec{k}_1',\mu_1',\vec{k}_2',\mu_2'\rangle=\\
&&=v_0\delta^3(\vec{v}- \vec{v}')\;f[\Delta m]\;\frac{(2\pi)^3}{
\sqrt{(\omega_1+\omega_2+\omega_3)^3(\omega_1'+\omega_2')^3}}\;
\langle k_1,\mu_1,k_2,\mu_2,k_3|\mathcal{H}_I(0)
|k_1',\mu_1',k_2',\mu_2'\rangle \label{kpseudo}
\end{eqnarray}
\end{widetext}
with $\omega_i=\sqrt{m_i^2+\vec{k}_i^2}$ denoting the particle
energies and $\Delta
m=\omega_1'+\omega_2'-\omega_1-\omega_2-\omega_3$.
$\mathcal{H}_I(0)$ is the pseudoscalar interaction Hamiltonian
density
\begin{equation}\label{pseudo}
-i\,g_{\mathrm{PS}}\;\bar{\psi}(0)\gamma_5\vec{\lambda}_F\psi(0)\cdot
\vec{\phi}(0)\;,
\end{equation}
evaluated at the space-time point
$x=0$. $\psi$ and $\phi$ are the fermion and boson fields,
$\vec{\lambda}_F$ the Gell-Mann flavor matrices, and $g_{\mathrm{PS}}$ is
the pseudoscalar quark-meson coupling constant.
 The kinematical factor in front of the matrix element
$\langle k_1,\mu_1,k_2,\mu_2,k_3|\mathcal{H}_I(0)
|k_1',\mu_1',k_2',\mu_2'\rangle$ has been chosen such that
$\langle k_1,\mu_1,k_2,\mu_2,k_3|$ and
$|k_1',\mu_1',k_2',\mu_2'\rangle$ can be taken as usual momentum
states (with $\vec{k}_1+\vec{k}_2+\vec{k}_3=0$ and
$\vec{k}_1'+\vec{k}_2'=0$).

We now turn to dynamical equations. The bound-state problem of
nonrelativistic quantum mechanics is usually reduced to the
stationary Schr\"odinger equation
\begin{equation}\label{sgl}
H |\Psi\rangle =|\Psi\rangle E\;.
\end{equation}
In non-relativistic quantum mechanics the Hamiltonian $H$ is the
only generator of the Galilei group that contains interactions,
the remaining nine generators are kinematical, i.\,e.~free
of interactions. In relativistic quantum mechanics one deals with
the Poincar\'{e} group instead and due to the different underlying
Lie algebra at least three generators contain interaction terms.
In the point form, as already mentioned above, interactions are
contained in all components of the four-momentum whereas the
Lorentz generators are kinematical. This means that one has to
solve the system of dynamical equations
\begin{equation}\label{indeppfsgl}
P^\mu\;|\Psi\rangle =p^\mu\;|\Psi\rangle \;
\end{equation}
instead of Eq.~(\ref{sgl}) to obtain the simultaneous eigenstates
of the components of the four-momentum operator  $P^\mu$. Within the
Bakamjian-Thomas framework the interaction dependence of the
four-momentum operator becomes particularly simple (cf.
Eq.~(\ref{intfourmom})) and Eq.~(\ref{indeppfsgl}) reduces to
\begin{equation}\label{eveq}
M\;|\Psi\rangle =m\;|\Psi\rangle \;.
\end{equation}
We note that the $v$ dependence of the total wave function of the
system $\Psi(v,\vec{k}_1,\mu_1,\ldots)=\langle
v,\vec{k}_1,\mu_1,\ldots|\Psi\rangle$ factors out after
projection of Eq.~(\ref{eveq}) onto velocity states.

Starting from the coupled-channel mass operator of
Eq.~(\ref{ccmassop}) and eliminating the $i+1$ particle channel
one arrives at an equation for only $i$ particles, with
one-particle exchange between any two of them. We write down the
eigenvalue equation for the mass operator of the two-channel
problem:
\begin{equation}\label{mateigeq}
\begin{pmatrix}\mathcal{D}_i^0&K^\dagger\\K&\mathcal{D}_{i+1}^0\end{pmatrix}
\begin{pmatrix}|\Psi_i\rangle\\|\Psi_{i+1}\rangle\end{pmatrix}=
m\begin{pmatrix}|\Psi_i\rangle\\|\Psi_{i+1}\rangle\end{pmatrix}\;,
\end{equation}
where $|\Psi_i\rangle$ and $|\Psi_{i+1}\rangle$ are states living
on the $i$- and $i+1$-particle subspaces, respectively.
$\mathcal{D}_i^0$ and $\mathcal{D}_{i+1}^0$ denote the free $i$-
and $i+1$-particle masses, $m$ is the mass eigenvalue of the
system, and $K$ is the vertex operator coupling the two channels.
From these two coupled equations for $|\Psi_i\rangle$ and
$|\Psi_{i+1}\rangle$ the latter can be eliminated to yield
\begin{equation}\label{opeigeq}
K^\dagger
(m-\mathcal{D}_{i+1}^0)^{-1}K|\Psi_i\rangle=(m-\mathcal{D}_i^0)|\Psi_i\rangle\;.
\end{equation}
Using velocity states, one can now turn this equation into an
integral equation, which is a generalized eigenvalue equation,
because $m$ also appears in the propagator
$(m-\mathcal{D}_{i+1}^0)^{-1}$. The operator $K^\dagger
(m-\mathcal{D}_{i+1}^0)^{-1}K$ acts as an optical (one-particle
exchange) potential. This optical potential contains, in
principle, also loop contributions in which the exchange particle
is reabsorbed by the emitting particle. But since we are
interested in studying relativistic few-body systems which
describe the dynamics of effective degrees of freedom we neglect
such contributions and assume that they can be absorbed in the
(renormalized) mass of the emitting particle.

\section{The Dynamical Equation}
\label{eigequ} So far we have not been very specific about the
system we want to investigate. Equation~(\ref{mateigeq}) is a
general mass eigenvalue equation for any coupled two-channel
problem. Up to this this point one could use this equation for
various different systems as described at the end of this paper.
In this section we apply it to a confined system of a constituent
quark and a constituent antiquark interacting via pseudoscalar
meson exchange, in order to give a description of vector mesons.
We make use of the velocity-state representation, so that the
Bakamjian-Thomas properties of the interacting mass operator can
be fully exploited. Taking only the one-meson exchange dynamics
into account, we could immediately start with Eq.~(\ref{opeigeq})
with $K$ defined according to Eqs.~(\ref{kpseudo}) and
(\ref{pseudo}).

Quarks, however, have not been observed as free particles in nature
and are therefore subject to a confining force.  
For a system of a constituent quark and a constituent antiquark
interacting via pseudoscalar meson exchange, $i$ in
Eq.~(\ref{opeigeq}) is equal to 2 and one has
\begin{equation}\label{theeq}
(\mathcal{D}_2^0-m)|\Psi_2\rangle=
K^\dagger(\mathcal{D}_3^0-m)^{-1}K|\Psi_2\rangle\;.
\end{equation}
The right-hand side now corresponds to an eigenvalue-dependent
pseudoscalar meson-exchange potential.  In order to introduce
confinement in this equation in as simple a manner as possible we
modify the relativistic kinetic energy terms $\mathcal{D}_{2}^0$ and
$\mathcal{D}_{3}^0$ to include harmonic oscillator confinement.  As
a next step the two-particle wave function is expanded in terms of
harmonic oscillator eigenfunctions, which will lead to a
discretization of the problem and allow us to apply straightforward
techniques for the numerical solution of the equation after a standard
partial wave analysis has been carried out.

Confinement is included in Eq.~(\ref{theeq}) in the diagonal terms in
such a way that the two quarks are confined in the two-particle
channel as well as in the three-particle channel, whereas the third
particle, the exchange boson, is free.  As we will argue in the
following this still provides the correct Lorentz-transformation
properties.  In the two-particle channel one can introduce confinement
``by hand'' by substituting the free two-particle mass operator by a
confinement one, say $\mathcal{D}^c_2$.  Such a confinement term a
priori does not need to be of harmonic oscillator type.  The
constraint that guarantees Lorentz invariance is that the confinement
operator must not depend on the overall velocity and must be a
rotational scalar.  In the three-particle channel only the two-quark
subsystem is confined, which is not at rest and therefore has to be
transformed to the correct frame.  Using velocity states as a basis,
all internal momenta (and also the angular momenta via the
corresponding $D$-functions, see Eq.~(\ref{veltrans})) are rotated by
the same rotation as an effect of a Lorentz transformation.  Since
only scalar products of internal momenta appear in the diagonal terms,
nothing changes by an overall rotation and also the three-particle
confinement mass operator has the correct Lorentz-transformation
properties.

To introduce confinement in the outlined manner one has to make
the following replacements in Eq.~(\ref{theeq})
\begin{eqnarray}
\mathcal{D}^0_2&\longrightarrow& \mathcal{D}^c_2\quad\mbox{and}\\
\mathcal{D}^0_3&\longrightarrow& \mathcal{D}^c_3\;.
\end{eqnarray}
The operators $\mathcal{D}^c_2$ and $\mathcal{D}^c_3$ are
explained in more detail in App.~\ref{defs}.  They are
essentially square roots of the usual Schr\"odinger operator for
the three-dimensional isotropic harmonic oscillator.  Appendix
\ref{defs} contains also the velocity-state representation of
these operators as well as our actual notation for the
eigenvalues.

\section{Solution Method and Numerics}
\label{numerics}
In the preceding chapter we have encountered an equation which has the
structure of an eigenvalue equation, but the eigenvalue also appears
in the optical potential.  Therefore one cannot directly employ
standard techniques for eigenvalue problems.  We use the following
approach:
\begin{itemize}

\item One sets the eigenvalue in the optical potential to some preset
value $m_{\mathrm{pre}}$ and treats $m_{\mathrm{pre}}$ as a parameter. 
In this way the equation becomes a linear eigenvalue equation and can
be solved using standard techniques to obtain eigenvalues
$\lambda_j(m_{\mathrm{pre}})$, $j=1,2,\ldots$, (which, of course,
depend on the value of $m_{\mathrm{pre}}$).  

\item For different preset values $m_{\mathrm{pre}}$ the resulting
spectra will be different.  

\item Interpolating the spectra for different $m_{\mathrm{pre}}$ leads
to continuous functions $\lambda_j(m_{\mathrm{pre}})$ of the preset
eigenvalue $m_{\mathrm{pre}}$.  The graphs of these functions exhibit
the $m_{\mathrm{pre}}$-dependence of the spectrum caused by the
$m_{\mathrm{pre}}$-dependence of the optical potential.

\item To find the positions of the generalized eigenvalues $M_j$,
$j=1,2,\ldots$, one has to solve the equations
\begin{equation}\label{rescond}
m=\mathrm{Re}(\lambda_j(m))\;.
\end{equation}
We will call Eq.~(\ref{rescond}) the ``resonance
condition''.

\item We take the real part of $\lambda_j(m)$, since the eigenvalues
are in general (at least above the thresholds for the production of
the various exchanged mesons) complex numbers.  This is justified,
because the shifts and widths resulting from the hyperfine interaction
are actually perturbations to the pure oscillator spectra.  The
imaginary part of $\lambda_j(M_j)$ corresponds to the width of the
(resonant) state with mass $M_j$, i.\,e.
\begin{equation}
\Gamma(M_j)=2|\mathrm{Im}(\lambda_j(M_j))|\;.\label{width}
\end{equation}

\end{itemize}
A procedure of this kind has, e.\,g., also been adopted in
Refs.~\cite{Bl87} and \cite{Bl90}.

Our generalized eigenvalue equation for a two-particle system
interacting via one-particle exchange without confinement is basically
an integral equation.  After introducing the confinement terms in the
diagonal parts of the coupled-channel mass operator of
Eq.~(\ref{ccmassop}), it is natural to use a harmonic-oscillator basis
for the quark-antiquark wave function.  By expressing the
$q$-$\bar{q}$ eigenstate $|\Psi_{2}\rangle$ in terms of the
eigenstates $|v,n,l,s,j,m_{j}\rangle$ of $\mathcal{D}^c_{2}$ and
taking only a finite number of basis states the eigenvalue equation
(\ref{theeq}) reduces to a system of coupled algebraic equations from
which the mass eigenvalues $m$ and the expansion coefficients
$A_{n,l,s,j,m_{j}}$ (see Eq.~(\ref{excoeff})) have to be determined. 
The matrix elements of the optical potential between the harmonic
oscillator eigenstates appear in this set of coupled equations.  These
are nine-dimensional integrals and involve sums over various quantum
numbers.  The integrations are done using standard Monte-Carlo
techniques.  Thereby the statistical errors have always been kept
smaller than one \mev, i.\,e.~smaller than about one per mille.  In the
course of the numerical calculation the Wigner rotations for the
three-particle intermediate-state wave functions (see
Eq.~(\ref{intstate})) have been neglected.  This approximation seems
to be justified by two observations.  On the one hand, the pair of
three-particle wave functions which shows up in the completeness
relation for the three-particle intermediate state must have similar
arguments to contribute substantially to the nine-dimensional
integral.  But this also means that the corresponding Wigner-rotations
approximately compensate each other.  On the other hand, it has been
observed in the investigation of baryon form factors within point-form
dynamics~\cite{Waetal01} that Wigner-rotation effects are of minor
importance.  The numerical effort, by the way, would substantially
increase if Wigner rotations were included in our calculations.

\section{Instantaneous Approximation}
\label{nonpot}
In order to study the effects of the exchange particle in flight as
compared to the standard instantaneous treatment of particle exchange
we perform a nonrelativistic reduction of the optical potential in the
point-form mass operator.  This is done via standard techniques and
goes along with an ``instantaneous approximation'' of the propagator
in the optical potential.  ``Instantaneous approximation''
means that the propagator denominator $(\mathcal{D}_{3}^c-m)$
is replaced by the energy of the exchanged meson
$\sqrt{\vec{q}^2+m_{\mathrm{Mes}}^2}$.  In the non-relativistic limit
the argument of the form factor reduces to the square of the
three-momentum of the exchanged meson.  Suppressing the flavor part of
the hyperfine interaction, one arrives at the well known form for
pseudoscalar meson-exchange potential (see, e.\,g., \cite{Maetal87})
\begin{equation}\label{vnr}
V_{\mathrm{NR}}(\vec{k}',\vec{k})=\frac{g_{\mathrm{ps}}^2}{4\pi}\;
\frac{f^2[\vec{q}^2]}{4m_1m_2}\;
\frac{(\vec{\sigma}_1\cdot\vec{q})(\vec{\sigma}_2\cdot\vec{q})}{
\vec{q}^2+m_{\mathrm{Mes}}^2}\;,
\end{equation}
where $\vec{q}$ is given by
\begin{equation}
\vec{q}=\vec{k}'-\vec{k}
\end{equation}
with $\vec{k}$ and $\vec{k}'$ representing c.m. momenta of the
incoming and outgoing quarks, respectively.

\section{Model Parameters}
\label{model}
We adopt the parameterization of the chiral constituent quark model of
Ref.~\cite{Gletal98b} for our actual calculation of the vector-meson
spectrum.  In the following the properties and parameters of this
model are briefly reviewed.  In a constituent-quark model one deals
with constituent quarks instead of current quarks; the constituent
mass is generated dynamically and is larger than the corresponding
current quark mass.
\[
m_u=m_d=340\;\mev\qquad\mbox{and}\qquad m_s=500\;\mev
\]
turn out to be appropriate mass values for light and strange
(constituent) quarks.  These numbers can already be obtained
approximately from simple quark-model arguments; recent lattice
calculations \cite{Aoetal99,Sketal01} also hint at these values.  The
constituent quarks (and antiquarks) are confined and interact in
addition via the exchange of the lightest pseudoscalar mesons which
are the Goldstone bosons associated with chiral symmetry breaking. 
The vertex describing this interaction is constructed from the
well-known pseudoscalar interaction Hamiltonian density given in
Eq.~(\ref{pseudo}).  The interaction vertex involves the pseudoscalar
coupling constant $g_{\mathrm{PS}}$ between constituent quark and
exchange meson and one cutoff parameter $\Lambda_i$,
$i=\pi,K,\eta,\eta'$, for each meson.  The parameters $\Lambda_i$
occur in the meson-(anti)quark-vertex form factors $f_i(\Delta m)$
(see Eqs.~(\ref{intmassop}) and (\ref{formf})) that we are using. 
Following Ref.~\cite{Gletal98b} we have taken
\begin{equation}
f_i(\Delta m)=\sqrt{\frac{\Lambda^2_i-m_i^2}{\Lambda^2_i-m_i^2+\Delta
m^2}}
\end{equation}
for the functional form of these vertex form factors. The cutoff
parameters $\Lambda_i$ are related by
\begin{eqnarray}\label{cutoff}
\Lambda_i=\Lambda_0+\kappa\;m_i\;,
\end{eqnarray}
with $\Lambda_0=566.33\,\mev$, $\kappa=0.81$, and $m_i$ being the mass
of the pseudoscalar meson of type $i$.  These vertex form factors go
to one when $\Delta m$ reaches zero\footnote{For $\Delta m=0$ we have
four-momentum conservation at the vertex with all three particles
being on-mass-shell.  For t-channel exchange of massive particles this
can, of course, only happen for unphysical momenta, but it is just the
kinematical situation (also in instant form) where the influence of
the vertex form factor is supposed to vanish and the coupling is
supposed to become point-like.} and they go to zero like $1/\Delta m$
for $\Delta m\rightarrow\infty$, leading to an additional $1/\Delta
m^2$-decay of the exchange potential.  In Ref.~\cite{Gletal98b} such a
kind of form factor serves to smear out the contact term which occurs
when~(\ref{vnr}) is transformed to configuration space.  The
coupling constant $g_8=g_{\mathrm{PS}}$ for the pseudoscalar octet can
be derived from the $N\!-\!\pi$ coupling constant via the
Goldberger-Treiman relation.  The value quoted by Glozman et
al.~\cite{Gletal98b} is $g_{\mathrm{PS}}^2/4\pi=0.67$.  Furthermore,
two different coupling constants are used for the pseudoscalar meson
octet and singlet, respectively.  The ratio of the singlet to octet
couplings taken in Ref.~\cite{Gletal98b} is $(g_0/g_8)^2=1.34$.
For our calculations the charge of the exchange particles is
irrelevant; therefore the (small) mass differences between differently
charged particles of the same sort, e.\,g.~the $\pi^\pm$ and the
$\pi^0$, are neglected.  The values used for the pseudoscalar meson
masses are basically the physical masses.  As in
Ref.~\cite{Gletal98b} we take
\begin{eqnarray*}
m_\pi=140\;\mev,\quad m_K=498\;\mev,\quad m_\eta=547\;\mev,\\
\quad
\mbox{and}\quad m_{\eta'}=958\;\mev\;.
\end{eqnarray*}
\begin{table}
\renewcommand{\arraystretch}{1.5}
\begin{tabular*}{\columnwidth}{c@{\extracolsep\fill}ccccccc}
\hline\hline
\multicolumn{3}{c}{Quark masses $[$MeV$]$}&&
\multicolumn{4}{c}{Meson masses $[$MeV$]$}\\
\multicolumn{2}{c}{$m_u$, $m_d$}&$m_s$&&$m_\pi$&$m_K$&
$m_\eta$&$m_{\eta'}$\\ \hline
\multicolumn{2}{c}{340}&
\parbox{0.8cm}{\centering 500}&&
\parbox{0.8cm}{\centering 139}&
\parbox{0.8cm}{\centering 494}&
\parbox{0.8cm}{\centering 547}&
\parbox{0.8cm}{\centering 958}\\ \hline\hline
\multicolumn{4}{c}{\phantom{MMM}Meson-quark coupling}&&
\multicolumn{3}{c}{Confinement}\\
$g^2_{8}/4\pi$&
\multicolumn{2}{c}{$(g_0/g_8)^2$}&
$\Lambda_0$ $[\mev]$&$\kappa$&
$a$ $[\mev]$&
\multicolumn{2}{c}{$V_0$ $[\gev^2]$}\\ \hline
0.67&
\multicolumn{2}{c}{1.34}&
566.33&0.81&
312&\multicolumn{2}{c}{-1.04115}\\\hline\hline
\end{tabular*}
\caption{Parameters for the point form description of vector mesons
within the chiral constituent quark model.  Apart from $a$ and $V_{0}$,
parameters are taken from 
Ref.~\cite{Gletal98b}}
\label{v1param}
\end{table}

Two more parameters come from the harmonic-oscil\-lator treatment of
the quark-antiquark confinement.  We denote the eigenvalues of
$\mathcal{D}_{2}^c$, i.\,e.~the square root of the harmonic-oscillator
eigenvalues, by
\begin{equation}
M_{nl}=\sqrt{8\;a^2(2n+l+\frac{3}{2})+V_0+4\bar{m}^2}\;,
\end{equation}
where $a$ is the oscillator parameter, $4\bar{m}^2$ contains the
rest masses of the quark and antiquark, and $V_0$ leads to an overall
shift of the spectrum (for details, see App.~\ref{defs}). Since
confinement is introduced in Ref.~\cite{Gletal98b} in a different and
not easily comparable way, $a$ and $V_0$ are free parameters. $a$ is
fixed in such a way that $M_{00}$ and $M_{10}$ agree with the masses
of the ground state and the first excited state of the $\varrho$
spectrum. Doing this we get
\[
a\approx 312\;\mev\;.
\]
This is a reasonable procedure, because the difference
$(M_{10}-M_{00})$ is nearly independent of the additional hyperfine
interaction.  This value for the oscillator parameter $a$ is kept fixed
throughout all calculations.  From the spectrum of the full
calculations including the hyperfine interaction, $V_0$ is fixed to
yield the $\varrho$ ground state at $770\;\mev$.  A suitable value for
$V_0$ is $V_0=-1.04115\;\gev^2$.  All parameters of the model are
summarized in Tab.~\ref{v1param}.

We have also done calculations without vertex form factors.  For this
purpose all parameters are kept the same, only $V_0$ had to be
adjusted to yield the $\varrho$ ground state at $770$ \mev.  One gets
a slightly different value, namely $V_0=-1.04385\;\gev^2$.  The
calculations within the instantaneous approximation were
performed with the same set of parameters as the corresponding full
calculations.
Finally we note that the $\omega$ and $\phi$ flavor wave functions
used in our calculations are the ones which correspond to ideal mixing
of the singlet and octet states of $SU(3)_F$.

\section{Results and Discussion}
\label{results}
At the beginning of this section we want to emphasize that our primary
goal is not an optimal description of the meson spectrum, but
rather to demonstrate with a simple model how the multichannel
formalism developed works and how it differs from the standard
instantaneous treatment of particle exchange.  In our calculations we
have concentrated on the lowest-lying negative-parity light and
strange vector mesons, i.\,e.  mesons with $J^P=1^-$ ($J$ being the
total angular momentum and $P$ the parity of the system).  This
implies that only $l=0$ and $l=2$ states of the harmonic-oscillator
basis can contribute to the $q$-$\bar{q}$ wave function.  Whereas $l$
is a good quantum number when taking only the confining interaction
into account, $l=0$ and $l=2$ contributions start to mix if the
hyperfine interaction is turned on.  The numerical analysis, however,
reveals that the $l=2$ contributions have practically no effect on the
absolute masses (less than or at most $1$~\mev, which is also the
upper limit for our numerical accuracy).  Even if compared to the
level shift caused by the hyperfine interaction the $l=2$
contributions are negligible with the exception of the two excited
states of the $\omega$.  For these states the $l=2$ contributions
amount to 11\% (first) and 18\% (second excited state) of the total
level shift.  In all other cases the contributions lack significance
since they are smaller than the required numerical accuracy.  As
already explained in Sec.~\ref{numerics} the solution of the full
coupled channel problem involves an expansion of the vector-meson wave
functions in terms of harmonic oscillator eigenfunctions.  It turns
out that already three basis states are enough to obtain convergent
results on the per mille level for the ground and the first two
excited states.  For the instantaneous approximation of the meson
exchange the convergence properties are worse.  One needs about two
times as many basis states as in the calculation with the full optical
potential to achieve the required accuracy.  It should also be
mentioned that at those places where the harmonic oscillator
eigenfunctions appear in completeness relations for intermediate
states (cf.  Eqs.~(\ref{2punit}) and (\ref{3punit})) the upper limits
for the main quantum number $n$ and the orbital angular momentum
quantum number $l$ have been taken to be the same as in the expansion
of the $q$-$\bar{q}$ wave function.  \input{spectr1}

The spectrum of the lowest-lying vector mesons is plotted in
Fig.~\ref{specfig}.  The comparison of the full calculation and the
pure confinement result shows that the hyperfine interaction due to
(dynamical) Goldstone-boson exchange can be considered as a
perturbation.  Therefore the qualitative features of the vector-meson
spectrum are in our model essentially determined by the confinement
potential.  It is thus not too surprising that only the masses of the
ground states and the first excites states are comparable to
experiment, whereas the predictions for the second excited states lie
already much too high.  To obtain also quantitative agreement with
experiment it would certainly be necessary to take a confinement
potential which is more sophisticated than our simple
harmonic-oscillator confinement.  A refined confinement potential
which is applicable in momentum-space calculations has, e.\,g.~been
suggested in Ref.~\cite{He94}.  But as we said already at the
beginning of this section, we rather want to study particle exchange
within a relativistic framework and the conclusions about the particle
exchange should not depend too much on the specific choice of the
additional confinement potential.

The biggest level shifts caused by the hyperfine interaction are
detected for the $\omega$ spectrum.  This observation can already be
anticipated from the fact that the flavor factor at the $\pi$-quark
vertex, the pion being also the lightest exchange particle, has its
maximum value for the $\omega$ meson.  The $\omega$ spectrum is thus
also the best place to study the features of our treatment of particle
exchange.  The differences between the full calculation and the
instantaneous approximation are indeed seen to be most prominent in
this case.  Whereas the usage of a static meson-exchange potential for
the hyperfine interaction leads to an unphysically large splitting of
the $\rho$ and $\omega$ ground states, their approximate degeneracy is
nearly preserved by our dynamical treatment of the Goldstone-boson
exchange.  The $\omega$ spectrum is obviously also most sensitive to
the choice of the meson-quark vertex form factor.  Comparing the
results for the standard parameterization of the vertex-form factors
(see Sec.~\ref{model}) with the outcome for point-like coupling,
i.\,e.~ the form factors set to one, a striking observation can be
made: whereas the instantaneous approximation depends very strongly on
the form factor only a mild dependence is seen for the full
calculation.  The reason for this discrepancy is the difference in the
propagators that make up the hyperfine interaction.  In the
instantaneous approximation it is the (non-relativistic) meson
propagator, in the full optical potential it is rather the propagator
of the intermediate $q$-$\bar{q}$-meson state.  The $q$-$\bar{q}$
system in the intermediate state is in addition subject to confinement
which acts as a natural cutoff and damps the dependence on the vertex
form factors.

Our approach does not only cover recoil effects in particle exchange,
it provides, in principle, also non-perturbative predictions for
vector-meson decay widths.  As soon as the mass of a vector meson
excitation becomes larger than the ground state energy of the
confinement potential plus the mass of an exchange meson the
corresponding channel opens and the pseudoscalar meson can also be
emitted leaving a lower lying vector meson.  Above such a decay
threshold the optical potential and thus the eigenvalues acquire an
imaginary part and the width for the decay of the vector-meson
resonance into the open two-particle channels can be calculated via
Eq.~(\ref{width}).  Within our simple two-channel model the decay
modes are restricted to $\rho$-$\pi$, $\omega$-$\pi$, and
$\rho$-$\eta$.  Among the resonances in Fig.~\ref{specfig} there is
only one prominent, the $\omega(1420)$ which decays into $\rho$-$\pi$
with a measured width of $174\pm 40$~MeV. The experimental information
on the other resonance widths for the strong decay into one of the above
mentioned two-particle channels is rather poor.  Only upper bounds,
which are of the order of MeV are given.  Our theoretical results are
all below 1 MeV, i.\,e.~below our calculational accuracy.  In the
outlook we will discuss possible improvements of our model which may
also lead to larger decay widths.  It seems, however, unlikely that
the huge decay width of the $\omega(1420)$ can be explained within a
simple two-channel approach.  We rather expect that other
mechanisms than those included so far, e.\,g.~a strong final-state 
interaction, have to be taken into account.

\section{Summary and Outlook}\label{concl}
We have presented a Poincar\'{e} invariant and Lorentz covariant
point-form approach to the dynamical treatment of particle exchange. 
We have worked within the Bakamjian-Thomas framework, which means that
the invariant mass operator takes over the role of the Hamiltonian in
non-relativistic quantum mechanics.  Operators and wave functions have
been defined with respect to a velocity-state basis.  Velocity states
are very natural and advantageous for treating relativistic few-body
systems within point-form dynamics.  The starting point of our approach to
particle exchange is a two-channel problem in which the $i$ and
$(i+1)$ particle channels are coupled via a vertex interaction which
was derived from a field theoretical Hamiltonian density such that the
resulting mass operator is of Bakamjian-Thomas type.  By reducing the
problem to a one-channel problem for the $i$-particle channel we have
ended up with an optical potential which describes the dynamics of the
particle exchange.  The corresponding eigenvalue problem, however, is
non-linear and has to be solved by appropriate means.  Since this
framework accounts for particle production it is able to provide
non-perturbative predictions for (partial) decay widths of resonances.

As a first application of the developed formalism we have investigated
vector mesons within the chiral constituent-quark model in which the
hyperfine interaction between the confined quark-antiquark pair is
mediated by Goldstone-boson exchange, i.\,e.~by the exchange of the
lightest pseudoscalar mesons.  With a simple harmonic-oscillator
confinement and a parameterization of the chiral constituent-quark
model that has already been successfully applied for the description of
baryon spectra we have found that the hyperfine interaction due to
Goldstone-boson exchange causes only small level shifts.  Thus it can
be considered as a perturbation of the confinement interaction and the
confinement potential essentially determines the properties of the
mass spectrum.  The comparison of the results for the full optical
potential and the standard instantaneous meson-exchange potential
revealed sizable differences, in particular for the $\omega$ spectrum. 
These differences are also reflected in the sensitivity to the
parameterization of the meson-quark vertex form factors.  Whereas the
full calculation depends only mildly on the choice of the vertex form
factors, the instantaneous approximation is extremely
sensitive to changes in the form factors.  Since the meson-quark
couplings and the exchange-meson masses are subject to physical
constraints, any reasonable parametrization of the Goldstone boson
exchange can thus be expected to provide similar results in the full
calculation.  Our predictions for vector-meson decay widths lie below
the demanded numerical accuracy and thus lack significance.

Our conclusions from the investigation of vector me\-sons are that a
proper relativistic treatment of particle exchange has to go beyond
the standard instantaneous approximation and must account for the
dynamical behavior of the exchange particle.  The predictions for the
vector meson spectrum could be improved with a refined confinement
interaction.  For a reasonable description of resonance widths it may
be necessary to extend the optical potential by loop contributions,
i.\,e.~contributions in which the emitted meson is again absorbed by the
emitting particle.  For the present calculation we have assumed that
such contributions go as self-energy contributions into the
constituent-quark masses.  But this is at most an approximation since
the (anti)quark in a loop is not free, but confined.  Loop
contributions have, e.\,g., also been seen to be important in the
semirelativistic treatment of the nucleon-nucleon system if one
reaches the pion-production threshold~\cite{Elster98}.  It will be
worthwhile and necessary to investigate their role in our
coupled-channel formalism.  This formalism should also be useful in
treating other relativistic few-body systems which interact via
particle exchange.  The positronium and hydrogen systems are presently
under investigation.  They are well studied within instant- and
front-form dynamics and would allow for a comparison of the different
approaches and forms of relativistic dynamics.

\begin{acknowledgments}
A.~K.~was supported by the Austrian Science Foundation FWF under
grant Nr.~P14794. He is also grateful to the Dekanat der
Naturwissenschaftlichen Fakult\"at der Universit\"at Graz for
financial support.
\end{acknowledgments}

\begin{appendix}
\begin{widetext}
\section{Normalizations and matrix elements}
\label{defs}
In this appendix we collect the most important definitions and
formulae used in the calculation.  We start with some definitions
concerning velocity states.  Consider a system with overall four
velocity $v$ consisting of $n$ (spin 1/2) fermions with masses $m_i$.
Their spins and momenta are uniquely specified by their spin
projections $\mu_i$ and momenta $\vec{k}_i$, $i=1,\ldots,n$ in the
overall rest frame of the system.  We also define
$\omega_i:=(m_i^2+\vec{k}_i^2)^{1/2}$.  Then the completeness relation
for the $n$-particle velocity states reads
\begin{eqnarray}\label{completerel}
\frac{1}{(2\pi)^{3n}}\sum_{\mu_1,\mu_2,\ldots,\mu_n}\int\frac{d^3v}{v_0}\left(
\prod_{i=1}^{n-1}d^3k_i\right)
\frac{(\sum_{i=1}^n\omega_i)^3}{\prod^{n}_{i=1}2\omega_i}
|v,\vec{k}_1,\mu_1,\vec{k}_2,
\mu_2,\ldots,\vec{k}_n,\mu_n\rangle\langle v,\vec{k}_1,\mu_1,
\vec{k}_2,\mu_2,\ldots,\vec{k}_n,\mu_n|=1\;;
\end{eqnarray}
the corresponding orthogonality relation is
\begin{eqnarray}\label{normrel}
\langle v,\vec{k}_1,\mu_1,\vec{k}_2,\mu_2,\ldots,\vec{k}_n,\mu_n|
v',\vec{k}_1',\mu_1',\ldots,\vec{k}_n',\mu_n'\rangle
=(2\pi)^{3n}\frac{\prod^{n}_{i=1}2\omega_i}{(\sum_{i=1}^n\omega_i)^3}\;
v_0\delta^3(\vec{v}- \vec{v}')
\prod_{i=1}^{n-1}\delta^3(\vec{k}_i-\vec{k}'_i)
\prod_{i=1}^{n}\delta_{\mu_i\mu'_i}\;.
\end{eqnarray}
The representation of the $n$-particle free mass operator in the
basis of $n$-particle velocity states is
\begin{eqnarray}
&&\langle v,\vec{k}_1,\mu_1,\ldots,\vec{k}_n,\mu_n|
\mathcal{D}_n^0|v',\vec{k}_1',\mu_1',\ldots,
\vec{k}_n',\mu_n'\rangle= \nonumber \\
&&=
(2\pi)^{3n}\frac{\prod^{n}_{i=1}2\omega_i}{(\sum_{i=1}^n\omega_i)^3}\;
v_0\delta^3(\vec{v}- \vec{v}')
\prod_{i=1}^{n-1}\delta^3(\vec{k}_i-\vec{k}'_i)
\prod_{i=1}^{n}\delta_{\mu_i\mu'_i} \quad\sum_{j=1}^n
\sqrt{m_j^2+\vec{k}_j^2}\;.
\end{eqnarray}
A state representing the confined quark-antiquark pair is labelled by
the overall velocity $v$ of the pair and the internal (oscillator)
quantum numbers $n$ and $l$, the total spin $s$, as well as the total
angular momentum $j$ and its projection $m_j$.  The completeness
relation for such states is
\begin{equation}
\frac{1}{(2\pi)^{3}}\sum_{l=0}^\infty\;\sum_{n=l}^\infty \;\sum_{s=0}^1\;
\sum_{j=|l-s|}^{l+s}\;
\sum_{m_{j}=-j}^j
\int\frac{d^3v}{v_0}\,\frac{M^2_{nl}}{2}\,|vnlsjm_j\rangle\langle
vnlsjm_j|=1\;,\label{2punit}
\end{equation}
where $M_{nl}^2$ are just the harmonic oscillator eigenvalues (see
Eq.~(\ref{hoev})) with main quantum number $n$ and orbital angular
momentum quantum number $l$.  The corresponding orthogonality relation
is
\begin{equation}
\langle v'n'l's'j'm_j'|vnlsjm_j\rangle=(2\pi)^{3}
\frac{2}{M^2_{nl}}\;v_0\delta^3(\vec{v}- \vec{v}')\;\delta_{n'n}
\delta_{l'l} \delta_{s's} \delta_{j'j} \delta_{m_j'm_j}\;.
\end{equation}
The mass operator $\mathcal{D}^c_2$ for the confined pair in the above basis
is
\begin{equation}
\langle v'n'l's'j'm_j'|\mathcal{D}^c_2|vnlsjm_j\rangle=(2\pi)^{3}
\frac{2}{M^2_{nl}}\;v_0\delta^3(\vec{v}- \vec{v}')\;\delta_{n'n}
\delta_{l'l} \delta_{s's} \delta_{j'j} \delta_{m_j'm_j}\;M_{nl}\;.
\end{equation}
Expansion coefficients $A_{nlsjm_{j}}$ of the $q$-$\bar{q}$ wave
function with respect to the harmonic-oscillator basis are defined by
\begin{equation}\label{excoeff}
\langle vnlsjm_j|\Psi\rangle=\langle vnlsjm_j| V,\Psi_{\mathrm{int}}
\rangle= (2\pi)^{\frac{3}{2}} \frac{\sqrt{2}}{M_{nl}}
\;v_0\delta^3(\vec{v}- \vec{V})\;A_{nlsjm_{j}}\;.
\end{equation}
For our problem of (negative parity) vector mesons $j=1$ and
$m_{j}=-1,0,1$ are fixed.  Furthermore, parity restricts spin and
orbital angular momentum to $s=1$ and $l=0,2$, so that the
coefficients $A_{nlsjm_{j}}$ depend de facto only on $n$ and $l$.  A
state describing a system of a quark-antiquark cluster and a free
pseudoscalar meson is labelled by the quantum numbers of the cluster,
the overall velocity $v$ and the relative momentum between the cluster
and the third particle $\kappa$.  The completeness and orthogonality
relations for such states are
\begin{equation}
\frac{1}{(2\pi)^{6}}
\sum_{l=0}^\infty\;
\sum_{n=l}^\infty\;
\sum_{s=0}^1\;
\sum_{j=|l-s|}^{l+s}\;
\sum_{m_{j}=-j}^j
 \int\frac{d^3v}{v_0}\, d^3\kappa
\,\frac{(\omega_{\mathrm{cl}}+ \omega_3)^3}{ 2\omega_{\mathrm{cl}}
2\omega_3}\,|v\kappa nlsjm_j\rangle\langle v\kappa nlsjm_j|=1
\label{3punit}\end{equation}
and
\begin{equation}
\langle v'\kappa' n'l's'j'm_j'|v\kappa nlsjm_j\rangle=(2\pi)^{6}
\frac{ 2\omega_{\mathrm{cl}} 2\omega_3}{(\omega_{\mathrm{cl}}+
\omega_3)^3}\;v_0\delta^3(\vec{v}- \vec{v}')\;
\delta^3(\vec{\kappa}- \vec{\kappa}')\;\delta_{n'n} \delta_{l'l}
\delta_{s's} \delta_{j'j} \delta_{m_j'm_j}\;,
\end{equation}
respectively. The factors in the Jacobian are defined by
$\omega_{\mathrm{cl}}=\sqrt{M_{nl}^2+\kappa^2}$ and
$\omega_3=\sqrt{m_3^2+\kappa^2}$. The mass operator $\mathcal{D}_3^c$ for
the confined pair and the free third particle in this basis takes
the form
\begin{equation}
\langle v'\kappa' n'l's'j'm_j'|\mathcal{D}_3^c |v\kappa nlsjm_j\rangle=
(2\pi)^{6} \frac{ 2\omega_{\mathrm{cl}}
2\omega_3}{(\omega_{\mathrm{cl}}+
\omega_3)^3}\;v_0\delta^3(\vec{v}- \vec{v}')\;
\delta^3(\vec{\kappa}- \vec{\kappa}')\;\delta_{n'n} \delta_{l'l}
\delta_{s's} \delta_{j'j} \delta_{m_j'm_j}\;
(\omega_{\mathrm{cl}}+ \omega_3)\;.
\end{equation}

The states defined above can be combined to yield the wave
function for the confined quark-antiquark pair. One has
\begin{eqnarray}\nonumber
&&\langle
\tilde{v}, \vec{k}_1,\vec{k}_2,\mu_1,\mu_2|vnlsjm_j\rangle=\\
&&=(2\pi)^{\frac{9}{2}}\;v_0\delta^3(\vec{\tilde{v}}- \vec{v})\;
\frac{\sqrt{2}}{M_{nl}}\;\sqrt{\frac{2\omega_1
2\omega_2}{(\omega_1+\omega_2)^3}}\,\sum_{m_l=-l}^l\;
\sum_{m_s=-s}^s C^{jm_j}_{lm_lsm_s}
C^{sm_s}_{\frac{1}{2}\mu_1\frac{1}{2}\mu_2}\;
u_{nl}(|\vec{k}_1|)\; Y_{lm_l}(\hat{k}_1)\;.
\end{eqnarray}
The functions $u_{nl}(k)$ are the well-known eigenfunctions of the
three-dimensional isotropic harmonic oscillator. Their explicit
form is
\begin{equation}
u_{nl}(k)=\frac{1}{\sqrt[4]{\pi}\;a^\frac{3}{2}}\;
\sqrt{\frac{2^{n+l+2}\;n!}{(2n+2l+1)!!}}\;
L_n^{l+\frac{1}{2}}\left(\frac{k^2}{a^2}\right)
\left(\frac{k}{a}\right)^{l} e^{-\frac{k^2}{2a^2}}\;,
\end{equation}
where $L_n^{l+\frac{1}{2}}$ is a generalized Laguerre polynomial.
The corresponding normalization integral is
\begin{equation}
\int\limits_0^\infty[u_{nl}(k)]^2\;k^2\;dk=1\;.
\end{equation}
The $Y_{lm_l}$ are the usual Spherical Harmonic functions.  For
further details see, e.\,g., Refs.~\cite{Me65} or \cite{El98}.  The
eigenvalues $M_{nl}$ of $D_{2}^c$ are the square root of
harmonic-oscillator eigenvalues, i.\,e.~
\begin{equation}\label{hoev}
M_{nl}=\sqrt{8\;a^2\left(2n+l+3/2\right)+V_0+4\bar{m}^2}\;.
\end{equation}
The well-known oscillator eigenvalues have been modified by adding an
overall spectral shift constant $V_0$ and an averaged rest-mass term
$\bar{m}^2$ to account for the different masses of the light and
strange constituent quarks.  The value of $\bar{m}$ in this expression
is easily determined for $\varrho$, $\phi$, and $\omega$, since the
masses of quark and antiquark are equal in these cases.  For the
$K^\ast$ we adopt an averaged mass squared of the form
\begin{equation}
\bar{m}^2=\frac{m_q^2+m_{\bar{q}}^2}{2}\;.
\end{equation}

For the system of the confined quark-antiquark pair and the third
free particle one can write
\begin{eqnarray}\nonumber
&&\langle\tilde{v}, \vec{k}_1, \mu_1,\vec{k}_2, \mu_2,\vec{k}_3
|v\kappa nlsjm_j\rangle=\\\nonumber
&&=(2\pi)^{\frac{15}{2}}\; v_0\delta^3(\vec{\tilde{v}}- \vec{v})\;
\delta^3(\vec{\kappa}-\vec{k}_3) \sqrt{\frac{2\omega_{\mathrm{cl}}
2\omega_3}{(\omega_{\mathrm{cl}}+\omega_3)^3}}\;
\sqrt{\frac{2\tilde{\omega}_1
2\tilde{\omega}_2}{2(\tilde{\omega}_1+ \tilde{\omega}_2)}}\;
\sqrt{\frac{2\omega_{12}2\omega_3}{(2\omega_{12}+2\omega_3)^3}}
\times\\\nonumber
&&\times\sum_{m_l=-l}^l\; \sum_{m_s=-s}^s\;
\sum_{\tilde{\mu}_1\tilde{\mu}_2=\pm\frac{1}{2}}
C^{jm_j}_{lm_lsm_s} C^{sm_s}_{\frac{1}{2}\tilde{\mu}_1\frac{1}{2}
\tilde{\mu}_2}\;
u_{nl}(|\vec{\tilde{k}}|)\; Y_{lm_l}(\hat{\tilde{k}})\;\times\\
&&\times\;D^{\frac{1}{2}}_{\mu_1\tilde{\mu}_1}
[B_{c}^{-1}(k_1/m_1)\;B_{c}(v_{12})\;B_{c}(\tilde{k}_1/m_1)]
\;D^{\frac{1}{2}}_{\mu_2\tilde{\mu}_2}[B_{c}^{-1}(k_2/m_2)
\;B_{c}(v_{12})\;B_{c}(\tilde{k}_2/m_2)] \; ,\label{intstate}
\end{eqnarray}
where $\tilde{k}=B_{c}^{-1}(v_{12})k_1$,
$v_{12}=
\begin{pmatrix} \sqrt{1+\frac{\kappa^2}{m_{12}^2}} \\
-\frac{\vec{\kappa}}{m_{12}}\end{pmatrix} $,
$\omega_{12}=\sqrt{m_{12}^2+\kappa^2}$,
$m_{12}=\tilde{\omega}_1+\tilde{\omega}_2$,
$\omega_{\mathrm{cl}}=\sqrt{m_{\mathrm{cl}}^2+\kappa^2}$, 
$D^\cdot_{\cdot\cdot}$ are Wigner $D$ functions, and
$\tilde{\omega}_i=\sqrt{\tilde{k}_i^2+m_i^2}$.

\end{widetext}

\end{appendix}

\bibliography{first}

\end{document}

%% file: spectr1.tex
\begin{figure*}[p]
\rotatebox{90}{
\begin{minipage}{23.5cm}
\beginpicture
\shaderectangleson
\setshadegrid span <0.4mm>
\setlength{\unitlength}{1mm}
\setcoordinatesystem units <0.8mm,0.08mm>
\thicklines
\setplotarea x from 0 to 280, y from 400 to
2000
\axis left ticks length <3mm> numbered from
600 to 2000 by 200
unlabeled length <1.5mm>  from 500 to 2000 by 100
unlabeled length <0.7mm> from 400 to 2000 by 50 /
\axis bottom 
 /
\put{$m$ [MeV]} at 0 2100
\put{\Large $\rho$} at 35 170
\put{\Large $\omega$} at 105 170
\put{\Large $\phi$} at 175 170
\put{\Large $K^\ast$} at 245 170
\put{EXP} at 10 320
\put{OSC} at 20 320
\put{PF} at 30 320
\put{IA} at 40 320
\put{NFF} at 50 320
\put{NIA} at 60 320
\put{EXP} at 80 320
\put{OSC} at 90 320
\put{PF} at 100 320
\put{IA} at 110 320
\put{NFF} at 120 320
\put{NIA} at 130 320
\put{EXP} at 150 320
\put{OSC} at 160 320
\put{PF} at 170 320
\put{IA} at 180 320
\put{NFF} at 190 320
\put{NIA} at 200 320
\put{EXP} at 220 320
\put{OSC} at 230 320
\put{PF} at 240 320
\put{IA} at 250 320
\put{NFF} at 260 320
\put{NIA} at 270 320
\setdashpattern <1mm, 0.4mm>
\data{10}{768}{769} \data{10}{1440}{1490}
\data{10}{1680}{1720}
\data{80}{782}{783} \data{80}{1388}{1450}
\data{80}{1625}{1673} 
\data{150}{1019}{1020} \data{150}{1660}{1700}
\data{220}{891}{892} \data{220}{1400}{1424}
\data{220}{1694}{1734}
%
\calc{30}{770}  \calc{30}{1467}
\calc{30}{1926}
\calc{100}{754}  \calc{100}{1456} \calc{100}{1917}
\calc{170}{1058}  \calc{170}{1636}
\calc{170}{2058} 
\calc{240}{927} \calc{240}{1554}
\calc{240}{1994}
%
\calc{50}{770}  \calc{50}{1467}
\calc{50}{1926}
\calc{120}{734}  \calc{120}{1447} \calc{120}{1910}
\calc{190}{1051}  \calc{190}{1632}
\calc{190}{2055} 
\calc{260}{926} \calc{260}{1554}
\calc{260}{1993}
%
\calc{20}{768} \calc{20}{1465}
\calc{20}{1925} 
\calc{90}{768} \calc{90}{1465}
\calc{90}{1925} 
\calc{160}{1062} \calc{160}{1638}
\calc{160}{2060} 
\calc{230}{926} \calc{230}{1554}
\calc{230}{1993} 
%
\calc{40}{776} \calc{40}{1470}
\calc{40}{1927} 
\calc{110}{708} \calc{110}{1407}
\calc{110}{1880} 
\calc{180}{1054} \calc{180}{1628}
\calc{180}{2050} 
\calc{250}{928} \calc{250}{1555}
\calc{250}{1994} 
%
\calc{60}{780} \calc{60}{1476}
\calc{60}{1932} 
\calc{130}{505} \calc{130}{1219}
\calc{130}{1774} 
\calc{200}{1039} \calc{200}{1599}
\calc{200}{2018} 
\calc{270}{927} \calc{270}{1554}
\calc{270}{1992} 
\endpicture \caption{The spectra for the lowest-lying light and
strange vector mesons.  The boxes in the columns labelled by
``EXP'' represent the experimental values with their 
uncertainties~\cite{PDG02}. 
The other columns give our numerical results for the pure
confinement interaction (``OSC''), the full calculation with
dynamical mesons exchange (``PF'') and the instantaneous approximation
to the meson exchange (``IA'').  Corresponding results with the
meson-quark vertex form factor set to one are labelled by ``NPF'' and
``NIA'', respectively.\label{specfig}}
\end{minipage}}
\end{figure*}